\documentclass[12pt]{article}
\usepackage{graphicx}
\usepackage{amsfonts}
\usepackage{amsmath}
\usepackage{amssymb}

\begin{document}

\title{L0+L1+L2 mixed optimization:  a geometric approach to seismic imaging and inversion using concepts in topology and semigroup}

\author{August Lau and Chuan Yin \\
        \\
        Apache Corporation \\
        2000 Post Oak Blvd., Houston, Texas 77056 \\
        \\
        Email contact: \texttt{chuan.yin@apachecorp.com}}
\date{June 29, 2010}

\maketitle

\begin{abstract}

The mathematical interpretation of L0, L1 and L2 is needed to understand how we should use these norms for optimization problems.   
The L0 norm is combinatorics which is counting certain properties of an object or an operator.  This is the least amplitude dependent norm since it is counted regardless of the magnitude.
The L1 norm could be interpreted as minimal geometric description.  It is somewhat sensitive to amplitude information.  In geophysics,  it has been used to edit outliers like spikes in seismic data.  This is a good application of L1 norm.
The L2 norm could be interpreted as the numerically simplest solution to fitting data with a differential equation.   It is very sensitive to amplitude information.  Previous application includes least square migration.
In this paper, we will show how to combine the usage of L0 and L1 and L2.  We will not be optimizing the 3 norms simultaneously but will go from one norm to the next norm to optimize the data before the final migration.
\end{abstract}

\section{Introduction}

We will start with a short overview of L0, L1 and L2.  

L0:  The Betti numbers of computational topology show that the more holes there are in the object,  the higher is the Betti number.  So L0 is used to minimize the Betti numbers. An example is a timeslice of the interference of many diffractions in a 3-D cube.  

L1:  Variation is defined to be the absolute value of the first derivative.   The L1 variational norm measures the "flat" part of the data and keep it unchanged.   A good example is a 1-D example in which there are sharp edges and flat dips.   The normal Fourier filter is L2 which will create sinusoidal shapes.   But the L1 variational norm will not disturb the flat events and will not create sinusoids. 

L2:  Straightforward application of L2 will create sinusoids since L2 has a Fourier interpretation. This is Parseval's theorem.  But if semigroups (e.g. diffusion maps) are  introduced to define the geometry of the data,  then the eigenfunctions will be used to generate their own geometric shapes independent of sinusoids.

The L0 norm is combinatorics which is counting certain properties of an object or an operator.  This is the least amplitude dependent norm since it is counted regardless of the magnitude.  If the absolute value of a number is raised to the power of 0,  then it is 1 or 0.  The SEG 2009 paper of "geometric simplicity" showed that the Betti numbers of computational topology could be viewed as combinatorics of counting holes of an object.   The Betti number is used as a zero/one measure to minimize L0 Betti norm.   Migration can only collapse diffractions of low Betti numbers.  Data should be conditioned to have low Betti numbers (low L0 norm) before migration.

The L1 norm could be interpreted as minimal geometric description.  It is somewhat sensitive to amplitude information.  In geophysics,  it has been used to edit outliers like spikes in seismic data.  This is a good application of L1 norm.  The reason is that L1 norm seeks the simplest geometry that fits the data.   Spikes create singularities which are not geometrically simple.   So they tend to be reduced significantly.  The SEG 2008 paper on "complex decomposition" with variational method showed that minimizing the L1 variational norm extracts the simple part of the data.  The simple part is comprehensible by numerical methods like wave equation.   The complement of the simple part is the complex part which is interpretable but the complex part is not useful for data fitting.  So L1 norm is used to extract simple geometry which is smooth and yet not sinusoidal.

The L2 norm could be interpreted as the numerically simplest solution to fitting data with a differential equation.   It is very sensitive to amplitude information.  Previous application includes least square migration.  Even though L2 norm is the most common application,  it is mostly  blind to geometry.  L2 applied literally is "bad" in terms of creating non-geologic shapes.   A good example is that +1 and -1 cancel each other.  But the coupling of +1 and -1 would be a valid solution to an L2 solution.  Since +1 and -1 are really two spikes,  it creates geometric spikes with opposite polarity (null space). If we make two smooth sinusoids opposite in polarity instead of +1 and -1,  they will also create abnormal geometry, i.e.,  non-geologic shapes.   In order to mitigate such non-geologic shapes,  the SEG 2009 paper of "diffusion semigroup" introduced L2 semigroup norm to cluster diffusion geometry so that sinusoids could not be an eigenfunction in general.

\section{Methodology}

Let us start with field records.  We recommend applying the L1 variational norm as the first step,  then the L0 Betti norm,  and finally the L2 semigroup norm.    

Applying L1 variational norm first serves the purpose of removing singular events like spikes and amplitudes which do not obey data smoothness.   Since the input data is unmigrated,  we assume that it will be beneficial to remove these singularities first.  After that,  there is a decision point as to applying L2 semigroup norm right away or compute L0 Betti norm next.    We will describe a generic processing sequence on a dataset to illustrate the process.

\section{Numerical examples}

Our first example is removing singular event from field record. Firgure 1(a) shows a zoom-in view on a portion of a field record that contains a "spike", or a singular event, that doest not obey smoothness. Figure 1(b) shows the result of applying variational norm L1 to remove the "spike" or singular event. 
\begin{figure}
\centering
  \includegraphics[width=3in]{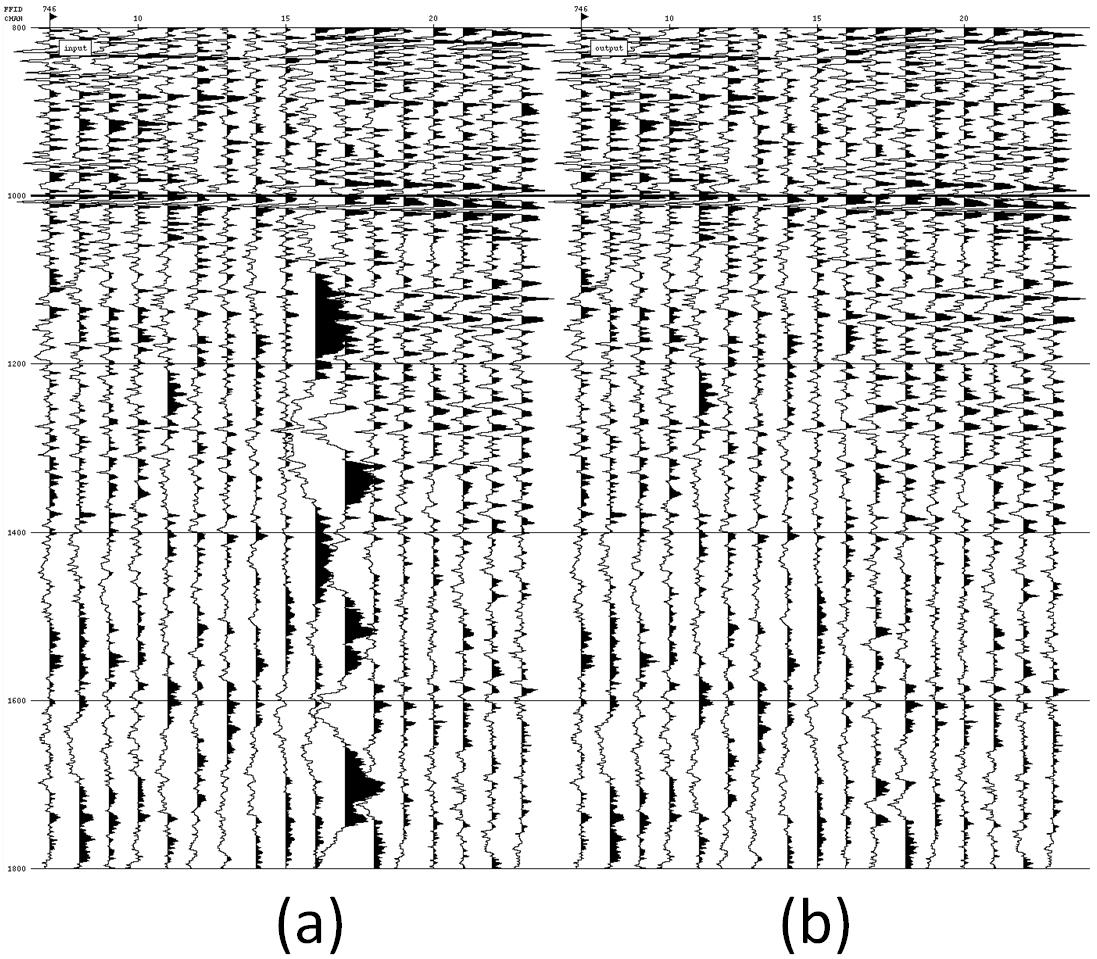}
\caption{(a) zoom-in plot of a field record with singular events; (b) Edited field record using L1 variational norm.}
\label{fig1}
\end{figure}
\begin{figure}
\centering
  \includegraphics[width=3in]{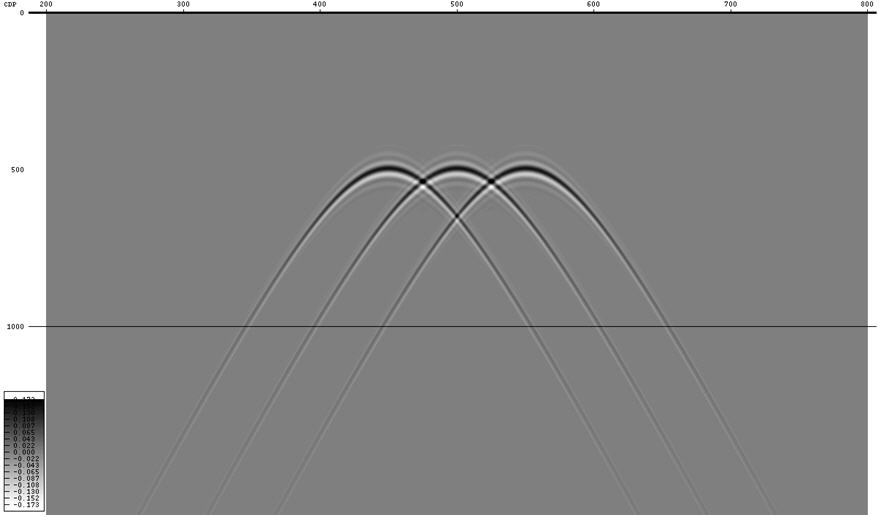}
\caption{A synthetic dataset containing three diffractors.}
\label{fig2}
\end{figure}
\begin{figure}
\centering
  \includegraphics[width=3in]{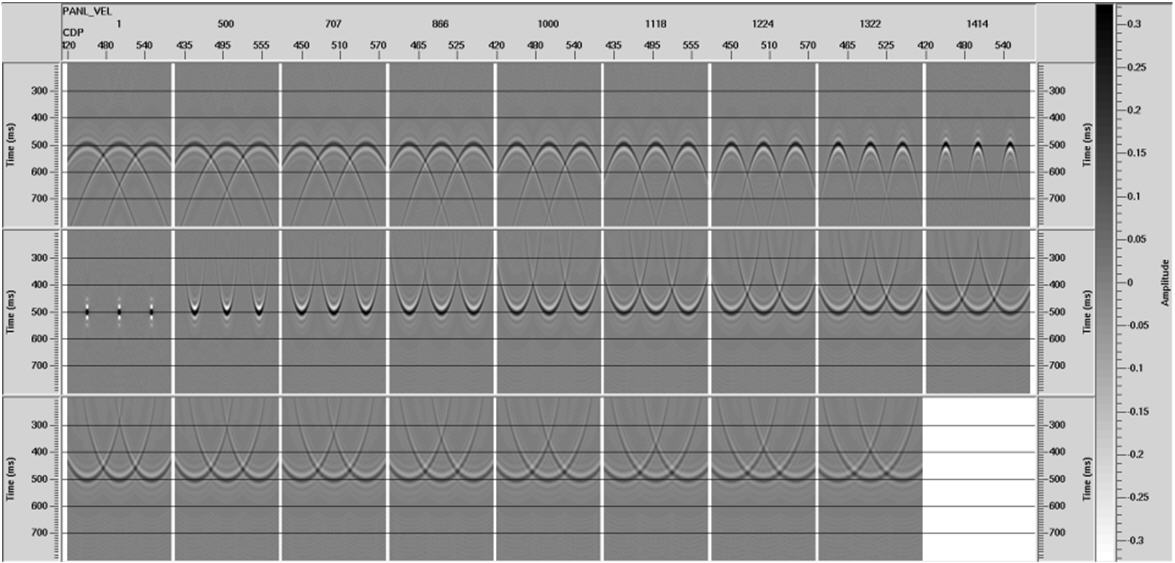}
\caption{Zoom-in view on the semigroup migration.}
\label{fig3}
\end{figure}

To demonstrate usage of L0 Betti norm for optimizing imaging, we first use a synthetic dataset containing three diffractors (Figure 2). We migrate the dataset with a semigroup migration. Figure 3 shows a zoom-in view on the migration. There are 26 panels, representing a sequence of $A^0 u$, $A^1 u$, $A^2 u$, $A^3 u$, ..., $A^{25} u$, where the migration operator $A = g(v^2)$ with $v = 500$ m/s, and $u$ is the input seismic (zero-offset). The sequence starts from top left to right; continue onto the next row left to right, then the bottom row left to right. There are theoretical advantages in using semigroup migration especially in the context of optimization. Since the concept of semigroup migration is outside the scope of this abstract, we will discuss it later in a seperate paper. 

As we can already observe, as the migration velocity increases and approaching the correct velocity (1500 m/s), interference pattern simplifies. And as the migration velocity further increases, i.e., above the true velocity, the pattern becomes more complex again. This observation can be further quantified by calculating the Betti numbers, B1 in this case, on each migrated image. Figure 4 shows such a calculation of B1 as a function of migration velocities. As expected, B1 reaches minimum when the migration velocity is equal to the true model velocity.

\begin{figure}
\centering
  \includegraphics[width=3in]{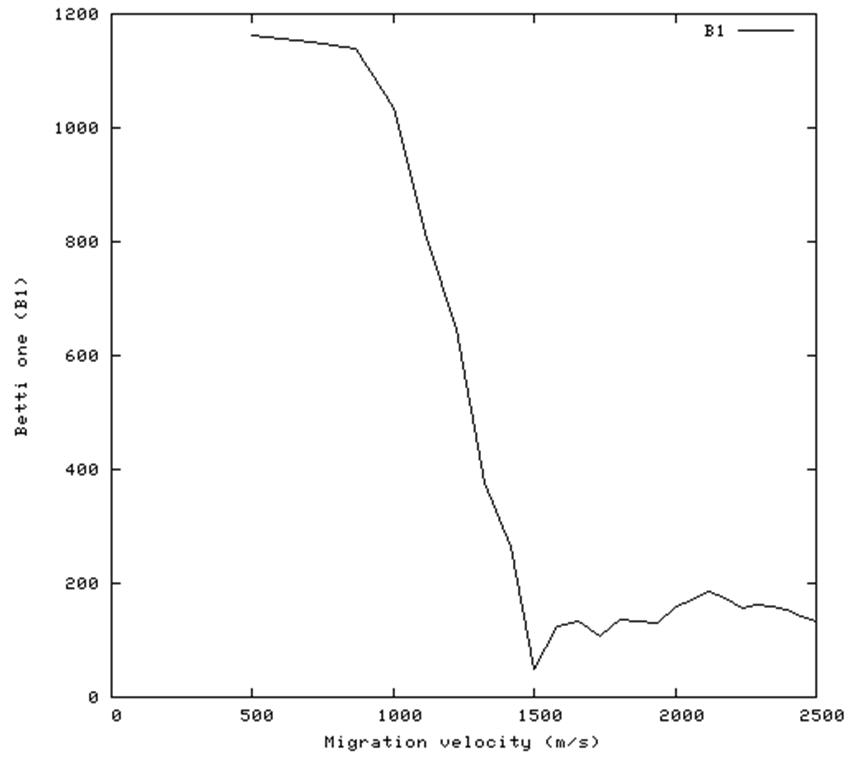}
\caption{Betti number B1 as a function of migration velocities for the synthetic dataset in Figures 2 and 3.}
\label{fig4}
\end{figure}

Although the above example is based on a synthetic record, it does demonstrate the importance of topological simplicity in optimization. Next we will show an example of applying L0 Betti norm to a real dataset. 

Firgure 5 shows a 2D stack section before migration. To help optimizing migration velocities, we again migrate the dataset with a semigroup migration. We then calculate Betti numbers to examine variation of topological complexity corresponding to different migration velocity. Figure 6 shows a zoom-in view on a portion of the dataset for different migration velocities. Note that the the interference pattern is less complex between velocities of 1500 m/s and 2000 m/s. Figure 7 quantifies the observation with a plot of Betti numbers versus migration velocity. It reaches minumum at a migration velocity of around 1800 m/s. 

\begin{figure}
\centering
  \includegraphics[width=3in]{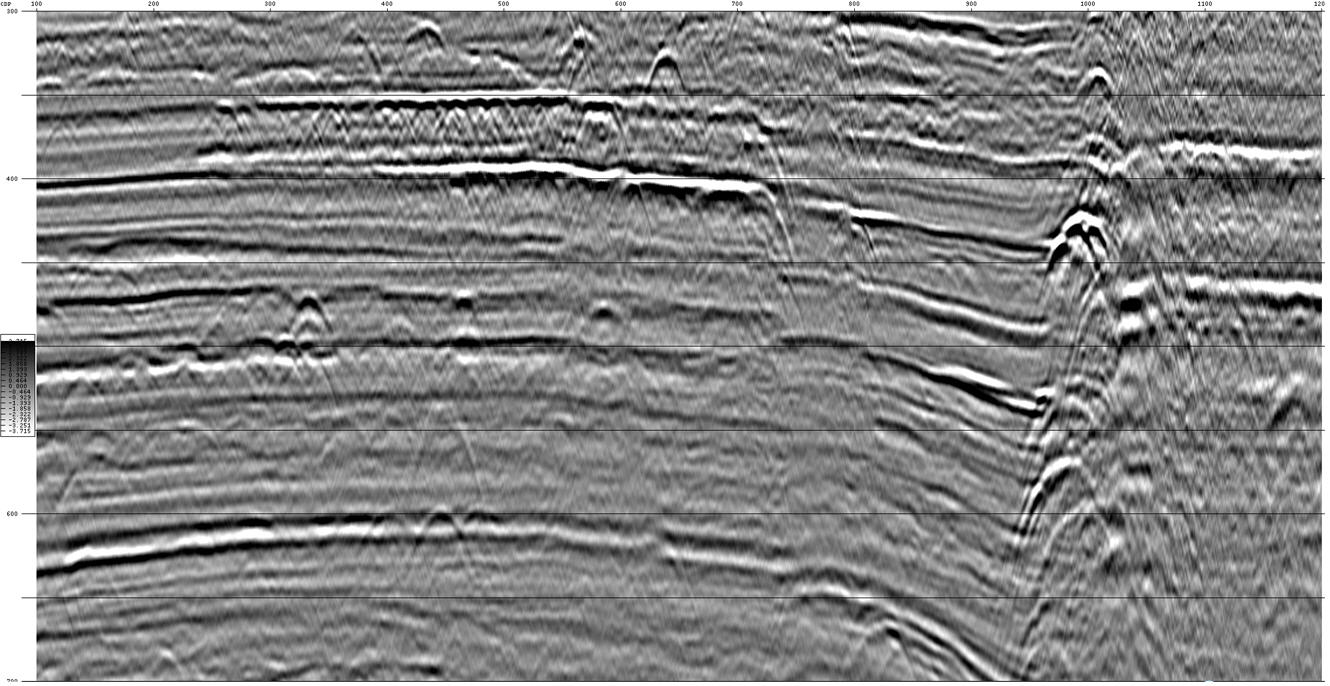}
\caption{A 2D stack section before migration.}
\label{fig5}
\end{figure}

\begin{figure}
\centering
  \includegraphics[width=3in]{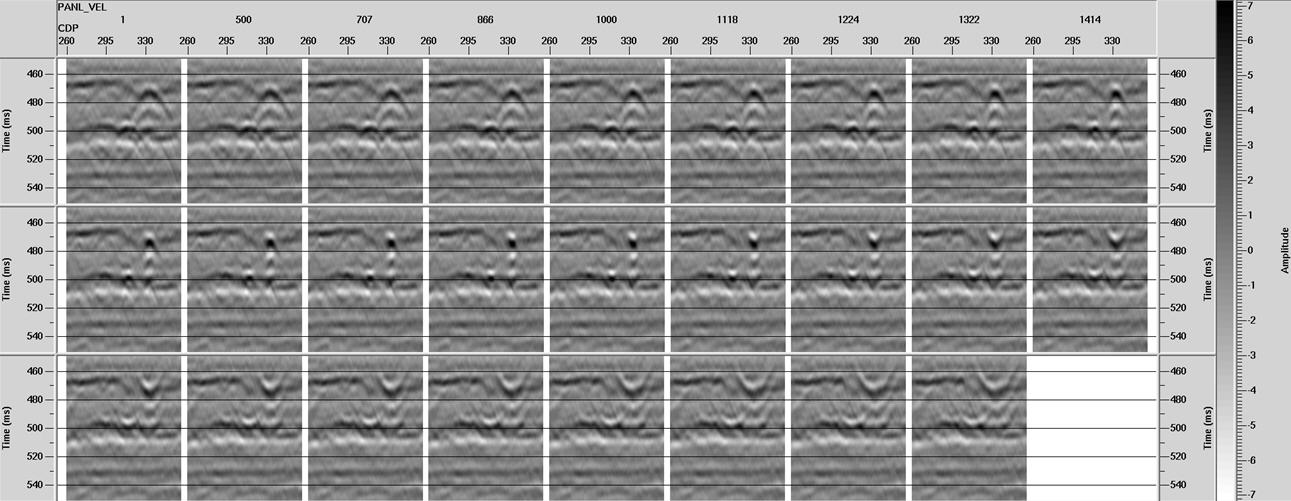}
\caption{Zoom-in view on a portion of the dataset from the semigroup migration.}
\label{fig6}
\end{figure}

\begin{figure}
\centering
  \includegraphics[width=3in]{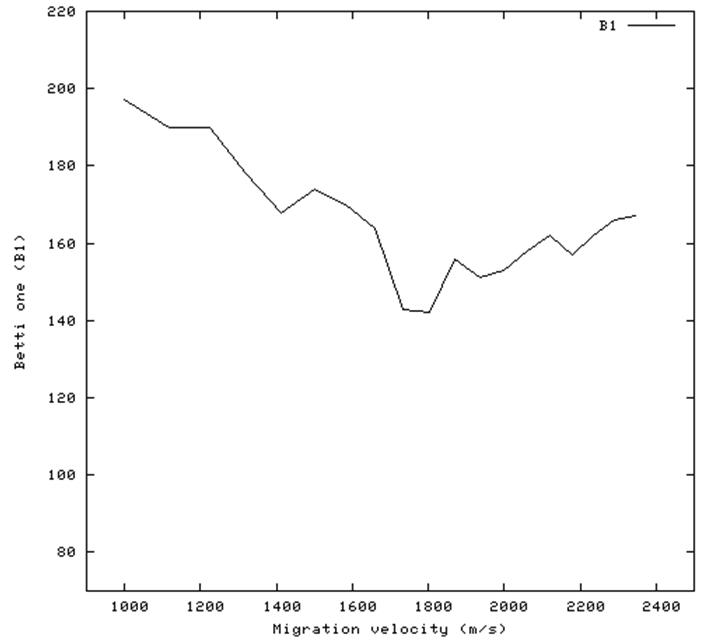}
\caption{Betti number B1 versus migration velocity, corresponding to Figure 6, that reaches minumum at a migration velocity of around 1800
 m/s.}
\label{fig7}
\end{figure}
\section{Conclusion} 

The 3 different norms of L0,  L1 and L2 serve different purposes.  The L0 is the least sensitive to amplitude since it counts the topology of the seismic by using Betti numbers.   This optimization is the most expensive but could be the most robust.   L1 is smoothing out singularities or replacing singularities like spikes with interpolated values.   This is less sensitive to amplitude compared to L2.   Finally,  L2 has the most efficient algorithm since it is based on quadratics.  However,  L2 does not comprehend intrinsic geometry of data.  Using L2 diffusion semigroup could help in defining data geometry without the detrimental effect of using predetermined geometry like Fourier or wavelet transforms.   By combining all three tools of L0 and L1 and L2,   we have a better opportunity to tackle complex optimization.

\bibliographystyle{plain}
\bibliography{L0L1L2}   

\end{document}